\documentclass[prd,twocolumn,preprintnumbers,amsmath,amssymb,nofootinbib]{revtex4}

\usepackage{amsmath}
\usepackage{graphicx}
\usepackage{xspace}
\usepackage{color}
\usepackage{url}
\usepackage[utf8]{inputenc}
\usepackage{epstopdf}
\usepackage{hyperref}
\usepackage{ragged2e}
\usepackage[T1]{fontenc}

\newcommand{\bea}{\begin{eqnarray}}
\newcommand{\eea}{\end{eqnarray}}

\newcommand{\eq}[1]{Eq.~\eqref{#1}}

\begin{document}
\preprint{CERN-TH-2021-012, PSI-PR-21-01, ZU-TH 01/21}

\title{Combined Constraints on First Generation Leptoquarks}

\author{Andreas Crivellin}
\email{andreas.crivellin@cern.ch}
\affiliation{CERN Theory Division, CH--1211 Geneva 23, Switzerland}
\affiliation{Physik-Institut, Universit\"at Z\"urich,
	Winterthurerstrasse 190, CH-8057 Z\"urich, Switzerland}
\affiliation{Paul Scherrer Institut, CH--5232 Villigen PSI, Switzerland}

\author{Dario M\"uller}
\email{dario.mueller@psi.ch}
\affiliation{Physik-Institut, Universit\"at Z\"urich,
	Winterthurerstrasse 190, CH-8057 Z\"urich, Switzerland}
\affiliation{Paul Scherrer Institut, CH--5232 Villigen PSI, Switzerland}

\author{Luc Schnell}
\email{luschnel@student.ethz.ch}
\affiliation{Departement Physik, ETH Zürich, Otto-Stern-Weg 1, CH-8093 Zürich, Switzerland}
\affiliation{D\'epartement de Physique, \'Ecole Polytechnique, Route de Saclay, FR-91128 Palaiseau Cedex, France}

\begin{abstract}
In this article we perform a combined analysis of low energy precision constraints and LHC searches for leptoquarks which couple to first generation fermions. Considering all ten leptoquark representations, five scalar and five vector ones, we study at the precision frontier the constraints from $K\to\pi\nu\nu$, $K\to\pi e^+e^-$, $K^0-\bar K^0$ and $D^0-\bar D^0$ mixing, as well as from experiments searching for parity violation (APV and QWEAK). We include LHC searches for $s$-channel single resonant production, pair production and Drell-Yan-like signatures of leptoquarks. Interestingly, we find that the recent non-resonant di-lepton analysis of ATLAS provides stronger bounds than the resonant searches recasted so far to constrain $t$-channel production of leptoquarks. Taking into account all these bounds, we observe that none of the leptoquark representations can address the so-called ``Cabibbo angle anomaly'' via a direct contribution to super-allowed beta decays.
\end{abstract}


\maketitle

\section{Introduction}

Leptoquarks (LQs) were first proposed in the context of the Pati-Salam model~\cite{Pati:1974yy} and $SU(5)$ Grand Unified theories~\cite{Georgi:1974sy,Dimopoulos:1980hn} but later on also postulated in
composite models with quark and lepton substructure~\cite{Schrempp:1984nj}, the strong
coupling version of the Standard Model (SM)~\cite{Abbott:1981re}, horizontal symmetry theories~\cite{Banks:1979fi}, extended technicolor~\cite{Lane:1993wz} as well as in $SO(10)$~\cite{Senjanovic:1982ex}, $SU(15)$~\cite{Frampton:1989fu}, superstring-inspired E6 models~\cite{Witten:1985xc} and the R-parity violating MSSM (see e.g. Ref.~\cite{Barbier:2004ez} for a review). With the HERA excess~\cite{Adloff:1997fg,Breitweg:1997ff} they came into the focus of the high energy community~\cite{Dreiner:1997cd,Kalinowski:1997fk,Kunszt:1997at,Altarelli:1997ce,Hewett:1997ce,Plehn:1997az} but after its disappearance the interest in LQ decreased.

Within recent years LQs experienced a revival, mainly due to the so-called ``flavor anomalies''. These are discrepancies between measurements and the SM predictions  which point towards lepton flavor universality (LFU) violating new physics (NP) in $R(D^{(*)})$~\cite{Lees:2012xj,Lees:2013uzd,Aaij:2015yra,Aaij:2017deq,Aaij:2017uff,Abdesselam:2019dgh}, $b\to s\ell^{+}\ell^{-}$~\cite{CMS:2014xfa,Aaij:2015oid,Abdesselam:2016llu,Aaij:2017vbb,Aaij:2019wad,Aaij:2020nrf} and in the anomalous magnetic moment (AMM) of the muon ($a_\mu$)~\cite{Bennett:2006fi}, with a significance of~$>\!3\,\sigma$~\cite{Amhis:2016xyh,Murgui:2019czp,Shi:2019gxi,Blanke:2019qrx,Kumbhakar:2019avh}, $>\!5\sigma$~\cite{Capdevila:2017bsm, Altmannshofer:2017yso,Alguero:2019ptt,Alok:2019ufo,Ciuchini:2019usw,Aebischer:2019mlg, Arbey:2019duh,Kumar:2019nfv} and \mbox{$>\!3\,\sigma$}~\cite{Aoyama:2020ynm}, respectively. In this context, it has been shown that LQs can explain $b\to s\ell^+\ell^-$ data~\cite{Alonso:2015sja, Calibbi:2015kma, Hiller:2016kry, Bhattacharya:2016mcc, Buttazzo:2017ixm, Barbieri:2015yvd, Barbieri:2016las, Calibbi:2017qbu, Crivellin:2017dsk, Bordone:2018nbg, Kumar:2018kmr, Crivellin:2018yvo, Crivellin:2019szf, Cornella:2019hct, Bordone:2019uzc, Bernigaud:2019bfy,Aebischer:2018acj,Fuentes-Martin:2019ign,Popov:2019tyc,Fajfer:2015ycq,  Blanke:2018sro,deMedeirosVarzielas:2019lgb,Varzielas:2015iva,Crivellin:2019dwb,Saad:2020ihm,Saad:2020ucl,Gherardi:2020qhc,DaRold:2020bib}, $R(D^{(*)})$~\cite{Alonso:2015sja, Calibbi:2015kma, Fajfer:2015ycq, Bhattacharya:2016mcc, Buttazzo:2017ixm, Barbieri:2015yvd, Barbieri:2016las, Calibbi:2017qbu, Bordone:2017bld, Bordone:2018nbg, Kumar:2018kmr, Biswas:2018snp, Crivellin:2018yvo, Blanke:2018sro, Heeck:2018ntp,deMedeirosVarzielas:2019lgb, Cornella:2019hct, Bordone:2019uzc,Sahoo:2015wya, Chen:2016dip, Dey:2017ede, Becirevic:2017jtw, Chauhan:2017ndd, Becirevic:2018afm, Popov:2019tyc,Fajfer:2012jt, Deshpande:2012rr, Freytsis:2015qca, Bauer:2015knc, Li:2016vvp, Zhu:2016xdg, Popov:2016fzr, Deshpand:2016cpw, Becirevic:2016oho, Cai:2017wry, Altmannshofer:2017poe, Kamali:2018fhr, Mandal:2018kau, Azatov:2018knx, Wei:2018vmk, Angelescu:2018tyl, Kim:2018oih, Aydemir:2019ynb, Crivellin:2019qnh, Yan:2019hpm,Crivellin:2017zlb, Marzocca:2018wcf, Bigaran:2019bqv,Crivellin:2019dwb,Saad:2020ihm,Dev:2020qet,Saad:2020ucl,Altmannshofer:2020axr,Fuentes-Martin:2020bnh,Gherardi:2020qhc,DaRold:2020bib} and/or $a_\mu$~\cite{Bauer:2015knc,Djouadi:1989md, Chakraverty:2001yg,Cheung:2001ip,Popov:2016fzr,Chen:2016dip,Biggio:2016wyy,Davidson:1993qk,Couture:1995he,Mahanta:2001yc,Queiroz:2014pra,ColuccioLeskow:2016dox,Chen:2017hir,Das:2016vkr,Crivellin:2017zlb,Cai:2017wry,Crivellin:2018qmi,Kowalska:2018ulj,Dorsner:2019itg,Crivellin:2019dwb,DelleRose:2020qak,Saad:2020ihm,Bigaran:2020jil,Dorsner:2020aaz,Fuentes-Martin:2020bnh,Gherardi:2020qhc,Babu:2020hun,Crivellin:2020tsz}, making them prime candidates for extending the SM with new particles.

Therefore, the investigation of LQ effects (in observables other than the flavor anomalies) is very well motivated. Complementary to direct LHC searches~\cite{Kramer:1997hh,Kramer:2004df,Faroughy:2016osc,Greljo:2017vvb, Dorsner:2017ufx, Cerri:2018ypt, Bandyopadhyay:2018syt, Hiller:2018wbv,Faber:2018afz,Schmaltz:2018nls,Chandak:2019iwj,Allanach:2019zfr, Buonocore:2020erb,Borschensky:2020hot}, leptonic observables~\cite{Crivellin:2020mjs} and oblique electroweak (EW) parameters as well as Higgs couplings to gauge bosons~\cite{Keith:1997fv,Dorsner:2016wpm,Bhaskar:2020kdr,Zhang:2019jwp,Gherardi:2020det,Crivellin:2020ukd} can be used to test LQs indirectly. Furthermore, if the LQs couple to first generation fermions particularly many low energy precision probes can be affected~\cite{Shanker:1981mj,Shanker:1982nd,Leurer:1993em,Leurer:1993qx,Davidson:1993qk}. Also beta decays can receive a tree-level effect from LQs which is interesting in the context of the so-called ``Cabibbo-Angle Anomaly''~\cite{Grossman:2019bzp,Seng:2020wjq}, where a (apparent) deficit in first row CKM unitarity can be reconciled via NP effects~\cite{Belfatto:2019swo,Coutinho:2019aiy,Crivellin:2020lzu,Capdevila:2020rrl,Crivellin:2020ebi,Kirk:2020wdk,Alok:2020jod,Crivellin:2020oup,Crivellin:2020klg}. Since a destructive effect w.r.t the purely left-handed SM amplitude is required by data, $SU(2)_L$ gauge invariance also leads to effects in rare Kaon decays and/or $D^0-\bar D^0$ in LQ models~\cite{Bobeth:2017ecx,Dorsner:2019vgp} which are complementary to LHC bounds. Therefore, it is interesting to investigate if it is possible to account for the Cabibbo angle anomaly once all other (relevant) available constraints are taken into account.

In this article we perform a complete analysis of all ten LQ representations, assuming only couplings to first-generation (weak-eigenstate) fermions to determine the combined allowed regions in parameter space. For this purpose, we define our setup and conventions in Sec.~\ref{sec:setup} and perform the matching on the relevant operators of the SM effective field theory (SMEFT). In Sec.~\ref{sec:observables} we calculate how the SMEFT coefficients are related to experimental constraints, perform the phenomenological analysis in Sec.~\ref{sec:pheno} and conclude in Sec.~\ref{conclusion}.

\section{Setup and Matching}
\label{sec:setup}

LQs have first been classified systematically in Ref.~\cite{Buchmuller:1986zs} into 10 possible representations under the SM gauge group: five scalar and five vector ones, as listed in Table~\ref{tab:LQ_representations}. The conventions are chosen such that the electric charge $Q$ is given by $Q=\frac{1}{2}Y+T_{3}$, where $Y$ is the hypercharge and $T_{3}$ the third component of the weak isospin. These representations allow for couplings to SM quarks and leptons as given in Table~\ref{tab:LQ_fermion_coupling}. Here we did not consider couplings to two quarks, which, together with the couplings in Table~\ref{tab:LQ_fermion_coupling}, would lead to proton decay. Note that such couplings can be avoided (to all orders in perturbation theory) by assigning baryon and/or lepton number to the LQs. In the following, we denote the LQ masses according to their representation and use small $m$ for the scalar LQs and capital $M$ for the vector LQs.

\begin{center}
	\begin{table}
		\begin{tabular}{c|ccccc|ccccc}
			Field & $\Phi_{1}$& $\tilde{\Phi}_{1}$ & $\Phi_{2}$ & $\tilde{\Phi}_{2}$ & $\Phi_{3}$ & $V_{1}^{}$ & $\tilde{V}_{1}^{}$ & $V_{2}^{}$ & $\tilde{V}_{2}^{}$ & $V_{3}^{}$\\
			\hline
			$SU(3)_{c}$ & 3 & 3 & 3 & 3 & 3 & 3 & 3 & 3 & 3 & 3\\
			$SU(2)_{L}$ & 1 & 1 & 2 & 2 & 3 & 1 & 1 & 2 & 2 & 3\\
			$U(1)_{Y}$ & $-\frac{2}{3}$ & $-\frac{8}{3}$ & $\frac{7}{3}$ & $\frac{1}{3}$ & $-\frac{2}{3}$ & $\frac{4}{3}$ & $\frac{10}{3}$ & $-\frac{5}{3}$ & $\frac{1}{3}$ & $\frac{4}{3}$
		\end{tabular}
		\caption{The ten possible representations of scalar and vector LQs under the SM gauge group.}
		\label{tab:LQ_representations}
	\end{table}
\end{center}

\subsection{Matching}

We now perform the tree-level matching of our ten LQ representations on $SU(2)_L$ gauge invariant dimension-six four-fermion operators using the basis of Ref.~\cite{Grzadkowski:2010es}
\begin{align}
\begin{aligned}
{\cal L}&{ = \sum {{C_i}} {O_i}}\,,\\
{O_{\ell q}^{(1)}}&{= [\bar Q{\gamma ^\mu }Q][\bar L{\gamma _\mu }L]}\,,\\
{O_{\ell q}^{(3)}}&{= [\bar Q  \tau^I {\gamma ^\mu }Q][\bar L \tau^I {\gamma _\mu }L]}\,,\\
{{O_{qe}}}&{= [\bar Q{\gamma ^\mu }Q][\bar e{\gamma _\mu }e]}\,,\\
{{O_{\ell u}}}&{= [\bar u{\gamma ^\mu }u][\bar L{\gamma _\mu }L]}\,,\\
{{O_{\ell d}}}&{ = [\bar d{\gamma ^\mu }d][\bar L{\gamma _\mu }L]}\,,\\
{{O_{eu}}}&{= [\bar u{\gamma ^\mu }u][\bar e{\gamma _\mu }e]}\,,\\
{{O_{ed}}}&= [\bar d{\gamma ^\mu }d][\bar e{\gamma _\mu }e]\,,
\end{aligned}
\end{align}
and find
\begin{widetext}
	\renewcommand{\arraystretch}{2.0}
\begin{equation}
\begin{array}{*{20}{c}}
{}&\vline& {C_{\ell q}^{(1)}}&{C_{\ell q}^{(3)}}&{{C_{qe}}}&{{C_{\ell u}}}&{{C_{\ell d}}}&{{C_{eu}}}&{{C_{ed}}}\\
\hline
{{\Phi _1}}&\vline& {\dfrac{{|\lambda _1^L{|^2}}}{{4m_1^2}}}&{ - \dfrac{{|\lambda _1^L{|^2}}}{{4m_1^2}}}&*&*&*&{\dfrac{{|\lambda _1^R{|^2}}}{{2m_1^2}}}&*\\
{{{\tilde \Phi }_1}}&\vline& *&*&*&*&*&*&{\dfrac{{|{{\tilde \lambda }_1}{|^2}}}{{2\tilde m_1^2}}}\\
{{\Phi _2}}&\vline& *&*&{ - \dfrac{{|\lambda _2^{LR}{|^2}}}{{2m_2^2}}}&{ - \dfrac{{|\lambda _2^{RL}{|^2}}}{{2m_2^2}}}&*&*&*\\
{{{\tilde \Phi }_2}}&\vline& *&*&*&*&{ - \dfrac{{|{{\tilde \lambda }_2}{|^2}}}{{2\tilde m_2^2}}}&*&*\\
{{\Phi _3}}&\vline& {\dfrac{{3|\lambda _3^2|}}{{4m_3^2}}}&{\dfrac{{|\lambda _3^2|}}{{4m_3^2}}}&*&*&*&*&*\\
{V_1 }&\vline& { - \dfrac{{|\kappa _1^L{|^2}}}{{2M_1^2}}}&{ - \dfrac{{|\kappa _1^L{|^2}}}{{2M_1^2}}}&*&*&*&*&{ - \dfrac{{|\kappa _1^R{|^2}}}{{M_1^2}}}\\
{\tilde V_1 }&\vline& *&*&*&*&*&{ - \dfrac{{|{{\tilde \kappa }_1}{|^2}}}{{\tilde M_1^2}}}&*\\
{V_2 }&\vline& *&*&{\dfrac{{|\kappa _2^{LR}{|^2}}}{{M_2^2}}}&*&{\dfrac{{|\kappa _2^{RL}{|^2}}}{{M_2^2}}}&*&*\\
{\tilde V_2 }&\vline& *&*&*&{\dfrac{{|{{\tilde \kappa }_2}{|^2}}}{{\tilde M_2^2}}}&*&*&*\\
{V_3 }&\vline& { - \dfrac{{3\left| {\kappa _3^2} \right|}}{{2M_3^2}}}&{\dfrac{{|\kappa _3^2|}}{{2M_3^2}}}&*&*&*&*&*
\end{array}
\end{equation}
\end{widetext}
in agreement with Ref.~\cite{Alonso:2015sja,Dorsner:2016wpm,Mandal:2019gff,Gherardi:2020det}

\begin{table}[t!]
	\renewcommand{\arraystretch}{1.7}
	\begin{tabular}{c|cc}
		{}& $L$&$e$\\
		\hline
		${\bar Q}$& ${\kappa_{1}^L{\gamma _\mu }V_1^{\mu} + \kappa _{3}{\gamma _\mu }\left( {\tau \cdot V_3^\mu } \right)}$&${\lambda_{2}^{LR}{\Phi _2}}$\\
		${\bar d}$& ${\tilde \lambda _{2}\tilde \Phi _2^Ti{\tau _2}}$&${\kappa_{1}^{R}{\gamma _\mu }V_1^{\mu}}$\\
		${\bar u}$& ${\lambda_{2}^{RL}\Phi _2^Ti{\tau _2}}$&${\tilde \kappa_1{\gamma _\mu }\tilde V_1^{\mu }}$\\
		${\bar Q_{}^c}$& ${\lambda_{3}i{\tau _2}{{\left( {\tau \cdot{\Phi _3}} \right)}^\dag } + \lambda_{1}^{L}i{\tau _2}\Phi _1^\dag }$&$\kappa_{2}^{LR}{\gamma _\mu }{V_2^{\mu \dag }}$\\
		${\bar d_{}^c}$& ${\kappa_{2}^{RL}{\gamma _\mu }V_2^{\mu \dag} }$&${\tilde \lambda_{1}\tilde \Phi _1^\dag }$\\
		${\bar u_{}^c}$& $\tilde{\kappa}_{2}{{\gamma _\mu }\tilde V_2^{\mu \dag }}$&${\lambda_{1}^{R}\Phi _1^\dag }$
	\end{tabular}
	\caption{Interaction terms of the LQ representations listed in Table~\ref{tab:LQ_representations}, where $Q$ and $L$ represent the left-handed quark and lepton $SU(2)_{L}$ doublets, $e$, $d$ and $u$ the right-handed $SU(2)_L$ singlets, the superscript $c$ stands for charge conjugation and $\tau_{i}$ are the Pauli matrices.}
	\label{tab:LQ_fermion_coupling}
\end{table}

For simplicity, we do not include flavor indices, since we will only consider couplings to first generation fermions (in the weak basis). Furthermore, we assume that $\Phi_1$, $\Phi_2$, $V_1$ and $V_2$ possess only one of the two possible couplings at the same time. Therefore, no scalar or tensor operators are generated, where the former ones are very stringently constrained from $\pi\to e\nu$.

Let us now consider the one-loop matching on four-quark operators~\cite{Aebischer:2020dsw} involving only left-handed fields:
	\begin{align}
	Q_{qq}^{(1)}&=\big[\bar{Q}\gamma^{\mu}Q\big]\big[\bar{Q}\gamma_{\mu}Q\big]\,,\\
	Q_{qq}^{(3)}&=\big[\bar{Q}{\tau^I}\gamma^{\mu}Q\big]\big[\bar{Q}{\tau}^I\gamma^{\mu}Q\big]\,,
	\end{align}
where the color indices are contracted within each bilinear and the Wilson coefficients are given by
\begin{subequations}
	\begin{align}
	\Phi_{1}:&&C_{qq}^{(1)}&=\frac{-|\lambda_{1}^{L}|^4}{256\pi^2 m_{1}^2}\,,&&C_{qq}^{(3)}=\frac{-|\lambda_{1}^{L}|^4}{256\pi^2 m_{1}^2}\,,\\
	\Phi_{2}:&&C_{qq}^{(1)}&=\frac{-|\lambda_{2}^{LR}|^4}{128\pi^2 m_{2}^2}\,,&&\\
	\Phi_{3}:&&C_{qq}^{(1)}&=\frac{-9|\lambda_{3}|^4}{256\pi^2 m_{3}^2}\,,&&C_{qq}^{(3)}=\frac{-|\lambda_{3}|^4}{256\pi^2 m_{3}^2}\,,\\
	V_{1}^{}:&&C_{qq}^{(1)}&=\frac{-|\kappa_{1}^{L}|^4}{32\pi^2 M_{1}^2}\,,&&\\
	V_{2}^{}:&&C_{qq}^{(1)}&=\frac{-|\kappa_{2}^{LR}|^4}{32\pi^2 M_{2}^2}\,,&&\\
	V_{3}^{}:&&C_{qq}^{(1)}&=\frac{-3|\kappa_{3}|^4}{32\pi^2 M_{3}^2}\,,&&C_{qq}^{(3)}=\frac{-|\kappa_{3}|^4}{16\pi^2 M_{3}^2}\,.
	\end{align}
	\label{CSMEFT}
\end{subequations}
Due to $SU(2)_L$, these operators will necessarily give rise to $K^{0}-\bar K^0$ and/or $D^{0}-\bar D^0$ mixing after electroweak symmetry breaking. For the vector LQs we calculated the diagrams in Feynman gauge, i.e. neglecting Goldstone contributions. In this way a finite result is obtained and the estimate is conservative in the sense that the NP contribution obtained is smaller than (the finite part of) the one in unitary gauge where large logarithms involving the cut-off appear~\cite{Barbieri:2016las}.

\subsection{Electroweak Symmetry Breaking}

For left-handed quarks ``first generation'' is only well defined in the interaction basis as after electroweak symmetry breaking non-diagonal mass matrices for the quarks are generated. In order to work in the physical basis with diagonal mass terms, we have to rotate the quark fields\footnote{The same is true for charged leptons. However, in the limit of vanishing neutrino masses all rotation necessary to diagonalize the charged lepton mass matrix are unphysical since they can be absorbed into a field redefinition.}
\begin{align}
\begin{aligned}
d_{L,f}\to U_{fi}^{d_L}d_{L,i}\,,\\
d_{R,f}\to U_{fi}^{d_R}d_{R,i}\,,\\
u_{L,f}\to U_{fi}^{u_L}u_{L,i}\,,\\
u_{R,f}\to U_{fi}^{u_R}u_{R,i}\,,
\label{Urotations}
\end{aligned}
\end{align}
with the unitary matrices $U^{u_{L,R}}$ and $U^{d_{L,R}}$. While the right-handed rotations can be absorbed by a re-definition of the couplings and are thus unphysical, the left-handed ones form the Cabibbo-Kobayashi-Maskawa (CKM) matrix
\begin{align}
V_{fi}\equiv U^{u_L*}_{jf}U^{d_L}_{ji}\,.
\label{eq:def_CKM_matrix}
\end{align}
As we want to study first generation LQs (defined in the weak basis), and flavor violating effects involving first and second quark generation quarks are most stringently constrained, we can focus on the $2\times 2$ sector which is related to the relatively large Cabibbo angle $\theta_c\approx0.22$. We can thus parameterize the matrices in \eq{Urotations} as
\begin{align}
\begin{aligned}
U_{}^{uL} &=
	\begin{pmatrix}
	\cos(\alpha) & \sin(\alpha)\\
	-\sin(\alpha) & \cos(\alpha)
	\end{pmatrix},\;\\
U_{}^{dL} &=
	\begin{pmatrix}
	 \cos(\beta) & \sin(\beta)\\
	 -\sin(\beta) & \cos(\beta)
\end{pmatrix}\,.
	\label{eq:UuL_UdL_matrices}
\end{aligned}
\end{align}
Using Eq.~\eqref{eq:def_CKM_matrix} this yields
\begin{subequations}
\begin{align}
	V &=
	\begin{pmatrix}
	\cos( {\beta  - \alpha }) & \sin({\beta  - \alpha })\\
	 -\sin({\beta  - \alpha }) & \cos( {\beta  - \alpha })
	\end{pmatrix}\\
	&\stackrel{!}{=}
	\begin{pmatrix}
	\cos(\theta_c) & \sin(\theta_c)\\
	-\sin(\theta_c)& \cos(\theta_c)
	\end{pmatrix}\,.
\end{align}
\end{subequations}
Hence, we can write
\begin{align}
U_{}^{uL}& =
	\begin{pmatrix}
	\cos( {\beta-{\theta _c} })&\sin( {\beta-{\theta _c}  })\\
	-\sin( {\beta-{\theta _c} }) & \cos( { \beta-{\theta _c} })
	\end{pmatrix}\,.
\end{align}
If $\beta=0$, we work in the so-called down basis where no CKM elements appear in flavor changing neutral currents (FCNCs) with down-type quarks. On the other hand, if we choose $\beta=\theta_{c}$, we work in the up basis in which down-type FCNCs are induced via CKM elements while up-type FCNCs are absent.

\section{Observables}
\label{sec:observables}

\subsection{Charged Semi-Leptonic Current}
We use the charged current effective Hamiltonian
\begin{align}
\mathcal{H}_{\text{eff}}^{\ell\nu}=\frac{4G_{F}}{\sqrt{2}}V_{jk}{\hat{C}}_{jk}^{e\nu}\big[\bar{u}_{j}\gamma^{\mu}P_{L}d_{k}\big]\big[{\bar{e}} \gamma_{\mu}P_{L}\nu_{e}\big]\,,
\end{align}
governing semi-leptonic transitions. {The coefficients $\hat{C}_{jk}^{e\nu} = C_{jk}^\text{SM} + C^{e\nu}_{jk}$ are the sum of the SM and LQ contribution}. The normalization is chosen such that we have in the SM 
\begin{align}
C_{jk}^\text{SM}=\delta_{jk}\,.
\end{align}
Integrating out the LQs, we obtain the following tree-level matching results
\begin{align}
\begin{aligned}
C^{e\nu}_{11}=\frac{-1}{\sqrt{2}G_{F}}\frac{c_{\beta}c_{\beta-\theta}}{V_{ud}}C_{\ell q}^{(3)}\,,\\
C^{e\nu}_{12}=\frac{{-}1}{\sqrt{2}G_{F}}\frac{s_{\beta}c_{\beta-\theta}}{V_{us}}C_{\ell q}^{(3)}\,,
\end{aligned}
\end{align}
where we abbreviated \mbox{$c_{\beta}\equiv\cos(\beta)$}, \mbox{$s_{\beta}=\sin(\beta)$}, \mbox{$c_{\beta-\theta}\equiv\cos(\beta-\theta_c)$} and $s_{\beta-\theta}\equiv\sin(\beta-\theta_c)$ and neglected effects related to third generation quarks and charm quarks, which would result in much weaker limits than the bounds to be discussed now.

The $d\to ue\bar{\nu}_{e}$ transitions contribute to beta decays where the measured CKM element $V_{ud}^{\beta}$ (extracted from experiment using the SM hypothesis) is related to the unitary CKM matrix $V_{ud}^{L}$ of the Lagrangian (including NP effects)
\begin{align}
V_{ud}^{\beta}=V_{ud}^{L}\big(1+C_{11}^{e\nu_e}\big)\,.
\end{align}
The element $V_{ud}^{L}$ can then be converted to $V_{us}^{L}$ applying unitarity
\begin{align}
\big|V_{us}^{L}\big|=\sqrt{1-\big|V_{ud}^{L}\big|^2-\big|V_{ub}^{L}\big|^2}\,.
\end{align}
 We find
\begin{align}
\begin{aligned}
V_{us}^L \approx V_{us}^\beta  + \frac{{|V_{ud}^\beta {|^2}}}{{|V_{us}^\beta {|^2}}}{\mkern 1mu} C_{11}^{e{\nu _e}}\,.
\end{aligned}
\end{align}
$V_{ub}^\beta$ is most precisely determined from super-allowed beta decays. Following Ref.~\cite{Crivellin:2020ebi} we have
\begin{align}
V_{us}^\beta =0.2281(7)\,,\;\;V_{us}^\beta|_{\text{NNC}} =0.2280(14)\,,
\label{Vusbeta}
\end{align}
where the latter value contains the  ``new nuclear corrections'' (NNCs) proposed by Refs.~\cite{Seng:2018qru,Gorchtein:2018fxl}. Since at the moment the issue of the NNCs is not settled, we will quote results for both determinations. This value of $V_{us}^\beta$ can now be compared to $V_{us}$ from two and three body kaon~\cite{Aoki:2019cca} and tau decays~\cite{Amhis:2019ckw}
\begin{align}
\begin{split}
V_{us}^{K_{\mu 3}}&=0.22345(67)\,,\;\;\; V_{us}^{K_{e 3}}=0.22320(61)\,,\\
V_{us}^{K_{\mu 2}}&=0.22534(42)\,,\;\;\;\;\;V_{us}^\tau = 0.2195(19)\,,
\end{split}
\label{VusKl3}
\end{align}
which are significantly lower\footnote{During finalization of this article, Ref.~\cite{Shiells:2020fqp} obtained a value of $\left|V_{u d}\right|^{2}=0.94805(26)$ which even slightly increases the disagreement with $V_{us}$.}. This disagreement constitutes the so-called Cabibbo angle anomaly.

Besides $\beta$-decays, tests of LFU in pion and Kaon decays, defined at the amplitude level and normalized to unity in the SM, result in
\begin{align}
\begin{aligned}
\frac{\pi\to \mu\nu}{\pi\to e\nu}&\approx 1-\dfrac{C_{11}^{e\nu_e}}{V_{ud}}\,,\\\frac{K\to(\pi)\mu\nu}{K\to(\pi) e\nu}&\approx 1-\frac{C_{12}}{V_{us}}\,.
\end{aligned}
\end{align}
This has to be compared to
\begin{eqnarray}
\begin{aligned}
\frac{K\to\pi\mu\nu}{K\to\pi e\nu}\bigg|_{\exp}&=1.0010\pm 0.0025\,,\\
\frac{K\to\mu\nu}{K\to e\nu}\bigg|_{\exp}&=0.9978(18)\,,\\
\frac{\pi\to\mu\nu}{\pi\to e\nu}\bigg|_{\exp}&=1.0010(9)\,,
\end{aligned}
\end{eqnarray}
from Ref.~\cite{Moulson:Amherst}, Refs.~\cite{Ambrosino:2009aa,Lazzeroni:2012cx,Tanabashi:2018oca} and Refs.~\cite{Aguilar-Arevalo:2015cdf,Czapek:1993kc,Britton:1992pg,Tanabashi:2018oca}, respectively. Numerically, $C_{11}^{e\nu_e}\approx -0.001$ would significantly improve the agreement with data. Note that effects in charged current $D$ decays are not very constraining~\cite{Dorsner:2009cu}.

\subsection{Tree-Level Neutral Current}
Chiral quark-electron interactions can be constrained from atomic parity violation experiments like APV~\cite{Wood:1997zq,Dzuba:2012kx} and from the weak charge of the proton as measured by QWEAK~\cite{Allison:2014tpu,Androic:2018kni}. The relevant effective Lagrangian reads
\begin{align}
\mathcal{L}_{\text{eff}}^{ee}=\frac{G_{F}}{\sqrt{2}}\sum_{q=u,d}{\hat{C}_{1q}}\big[\bar{q}\gamma^{\mu}q\big]\big[\bar{e}\gamma_{\mu}\gamma_{5}e\big]\,,
\end{align}
{where $\hat{C}_{1q} = C_{1q}^{\text{SM}} + C_{1q}$ } with $C_{1u}^{\text{SM}}=-0.1887$ and $C_{1d}^{\text{SM}}=0.3419$. Again we can express the Wilson coefficients $C_{1q}$ in terms of the SMEFT matching coefficients
\begin{align}
C_{1u}&=\frac{-\sqrt{2}}{4G_{F}}\Big(c_{\beta-\theta}^2\big(C_{\ell q}^{(1)}-C_{\ell q}^{(3)}-C_{qe}\big)+C_{\ell u}-C_{eu}\Big)\,,\nonumber\\
C_{1d}&=\frac{-\sqrt{2}}{4G_{F}}\Big(c_{\beta}^2\big(C_{\ell q}^{(1)}+C_{\ell q}^{(3)}-C_{qe}\big)+C_{\ell d}-C_{ed}\Big)\,.
\end{align}
This has to be compared to~\cite{Zyla:2020zbs}
\begin{align}
Q_{W}(p)&=-2\left(2 {\hat{C}}_{1 u}+{\hat{C}}_{1 d}\right)=0.0719 \pm 0.0045\,,\\
Q_{W}\left({Cs}^{133}\right)&=-2\left(188 {\hat{C}}_{1 u}+211 {\hat{C}}_{1 d}\right)=-72.{82} \pm 0.42\,.\nonumber
\end{align}
For our numerical analysis we combine these constraints in a $\chi^2$ fit with one degree of freedom since each LQ representation predicts a single direction in $C_{1 u}-C_{1 d}$ space.

If we are not exactly aligned to the down basis (i.e. $\beta\neq0$), some representations generate $s\to d e^{+}e^{-}$ transitions which result in LFUV in $K\to\pi \mu^{+}\mu^{-}/K\to\pi e^{+}e^{-}$. With the current experimental constraints~\cite{Batley:2009aa,Appel:1999yq,Batley:2003mu} we find according to Ref.~\cite{Crivellin:2016vjc}
\begin{align}{
{{s_\beta }{c_\beta }\left( {C_{lq}^{(1)} + C_{lq}^{(3)} + {C_{qe}}} \right) = \dfrac{{0.0012 \pm 0.0046}}{{{\rm{Te}}{{\rm{V}}^2}}}}}
\end{align}
from $K^+\to\pi^+ \mu^{+}\mu^{-}/K^+\to\pi^+e^{+}e^{-}$. Similar test of LFU in $D$ decays are not constraining~\cite{Bause:2019vpr}.

Similarly, if the LQ representation couples left-handed down quarks to neutrinos, effects in $K\to\pi\nu\nu$ are generated for $\beta\neq0$. Here the the charged mode~\cite{Artamonov:2008qb}
\begin{align}
{\text{Br}}\big[K^{+}\to\pi^{+}\nu\bar{\nu}\big]=\big(1.73\substack{+1.15\\-1.05}\big)\times 10^{-10}\,,
\end{align}
provides better constraints and using the results of Ref.~\cite{Buras:2004qb,Buras:2015qea} we find
\begin{align}
&{\rm{Br}}\left[ {{K^ \pm } \to {\pi ^ \pm }\nu \bar \nu } \right] = \frac{1}{3}\left( {1 + {\Delta _{EM}}} \right){\eta _\pm }\times \nonumber\\
&\sum\limits_{f,i = 1}^3 \bigg[\frac{{\mathop{\rm Im}\nolimits} {{\big[ {{\lambda _t}\tilde X_{\nu}^{fi}} \big]^2}}}{\lambda ^{{5}}}+\! \bigg( \frac{{\mathop{\rm Re}\nolimits} \big[ {{\lambda _c}} \big]}{\lambda }{P_c}{\delta _{fi}} + \frac{{\mathop{\rm Re}\nolimits} \big[ {{\lambda _t}\tilde X_{\nu}^{fi}} \big]}{\lambda ^5} \bigg)^{\!2}\bigg]\,,
\end{align}
with $\lambda_{q}=V_{qs}^{*}V_{qd}$ and
\begin{align}
\begin{aligned}
&\tilde X_{\nu}^{fi} = X_{\nu}^{{\rm{SM}},fi} - {s_W^2}{C_{\nu}^{fi}}\,,\\
&X_{L}^{{\rm{SM}},fi} = \big( {1.481 \pm 0.005 \pm 0.008} \big){\delta _{fi}}\,,\\
&{P_c} = 0.404 \pm 0.024\,,~~~~{\Delta _{EM}} =  - 0.003\,,\\
&{\eta _\pm } = \big( {5.173 \pm 0.025} \big)\times{10^{ - 11}}{\bigg[ {\frac{\lambda }{{0.225}}} \bigg]^8}\,.
\end{aligned}
\end{align}
The LQ effects can again be expressed in the compact form
\begin{align}
C_{\nu}^{fi}&=\frac{-\pi s_{\beta}c_{\beta}}{\sqrt{2}G_{F}\alpha V_{ts}^{*}V_{td}}\Big(C_{\ell q}^{(1)}-C_{\ell q}^{(3)}\Big)\delta_{f1}\delta_{i1}\,,
\end{align}
by using \eq{CSMEFT}. Again, the analogous $D$ decays cannot complete in precision~\cite{Bause:2020xzj} and the loop-induced effects in $D^{0}-\bar{D}^{0}$ turn out to be more relevant.

\begin{figure*}
	\begin{center}
		\includegraphics[width=0.99\textwidth]{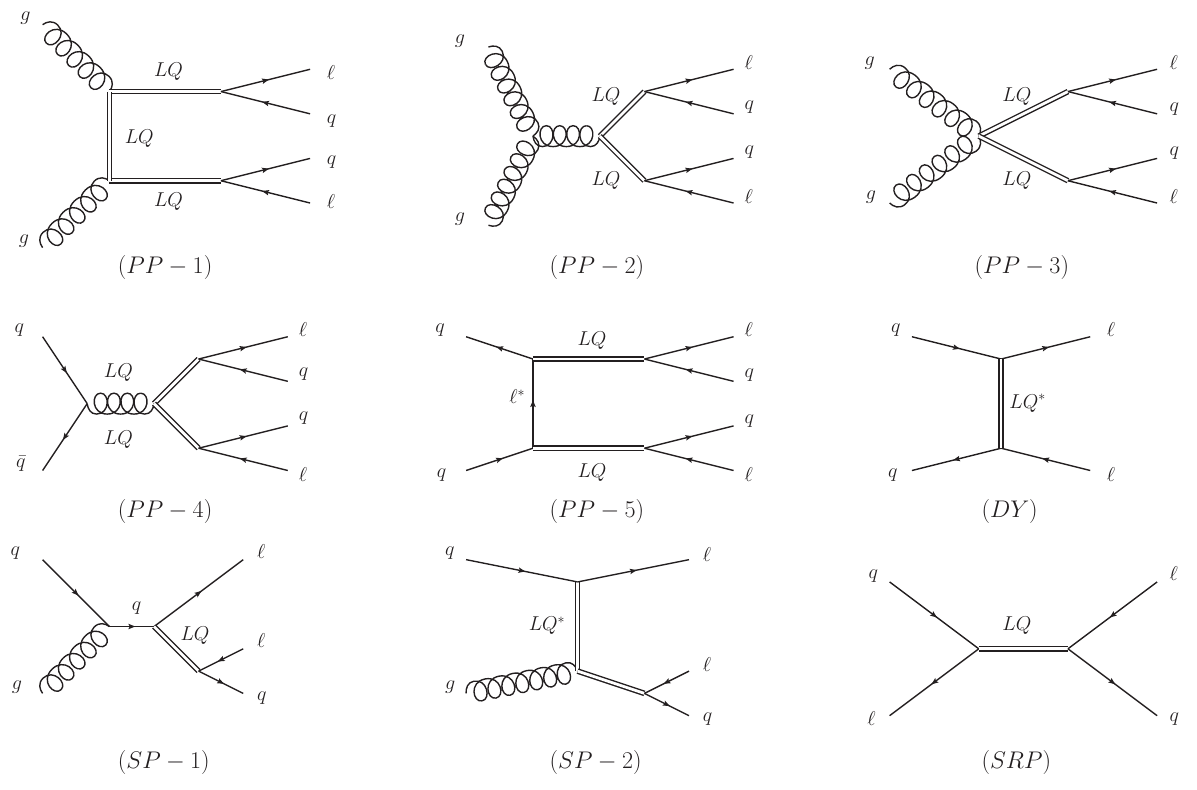}
	\end{center}
	\caption{Feynman diagrams showing the different search channels for LQs at the LHC.}
	\label{fig:FeynmanDiagrams}
\end{figure*}

\subsection{$D^{0}-\bar{D}^{0}$ and $K^{0}-\bar{K}^{0}$ Mixing}

Finally, if a LQ representation couples to left-handed quarks with $\beta\neq \theta_c$ $(\beta\neq 0)$ FCNC in $D^{0}-\bar{D}^{0}$ ($K^{0}-\bar{K}^{0}$) mixing is generated. We use
\begin{align}
\begin{aligned}
\mathcal{H}_{\text{eff}}^{D\bar{D}}&=C_{1}^D\big[\bar{u}_{\alpha}\gamma^{\mu}P_{L}c_{\alpha}\big]\big[\bar{u}_{\beta}\gamma_{\mu}P_{L}c_{\beta}\big]\,,\\
\mathcal{H}_{\text{eff}}^{K\bar{K}}&=C_{1}^K\big[\bar{d}_{\alpha}\gamma^{\mu}P_{L}s_{\alpha}\big]\big[\bar{d}_{\beta}\gamma_{\mu}P_{L}s_{\beta}\big]
\end{aligned}
\end{align}
{to parametrize NP contributions} and we obtain
\begin{align}
C_{1}^D&=-s_{\beta-\theta}^2c_{\beta-\theta}^2\Big(C_{qq}^{(1)}+C_{qq}^{(3)}\Big)\\
C_{1}^K&=-s_{\beta}^2c_{\beta}^2\Big(C_{qq}^{(1)}+C_{qq}^{(3)}\Big)
\end{align}
The limits on the coefficients are~\cite{Bona:2017gut}
\begin{align}
{
\begin{aligned}
|\text{Re}\big[C_{1}^{D}\big]|&\lesssim 3\times \frac{10^{-7}}{\text{TeV}^{2}}\,\\
|\text{Re}\big[C_{1}^{K}\big]|&\lesssim 1.3\times \frac{10^{-7}}{\text{TeV}^{2}}\,.
\end{aligned}}
\end{align}
Since the SM contribution cannot be reliably calculated in case of $D^{0}-\bar{D}^{0}$ mixing, we assumed that the NP contribution should not generate more than the whole measured mass difference to obtain this bound.

\subsection{LHC Bounds}
One can search for signals of LQs at the LHC generated via
\begin{itemize}
	\item Pair production (PP): $qq (gg)\to 2{\rm LQ}\to qq\ell\ell$
	\item Single production (SP): $qg\to {\rm LQ}\to \ell\ell q$
	\item Single resonant production (SRP): $\ell q\to {\rm LQ}\to \ell q$
	\item Drell-Yan (DY): $pp\to {\rm LQ^*}\to \ell\ell$
\end{itemize}
as depicted in Fig.~\ref{fig:FeynmanDiagrams}.

For first generation LQs, PP sets coupling independent limits on their masses. Here we use the bounds for the neutrino and charged lepton channels of Ref.~\cite{Sirunyan:2018kzh} and Ref.~\cite{Sirunyan:2018btu}, respectively. Note that the interactions of gluons with vector LQs depend on the nature of the LQ, i.e. whether it is a massive Proca field or a massive gauge boson~\cite{Blumlein:1996qp}. We choose the latter case (corresponding to \mbox{$\kappa_G=0$}) and rescaled the experimental bounds on the masses of Refs.~\cite{Sirunyan:2018kzh,Sirunyan:2018btu} by a constant factor $\approx 1.3$ derived from Ref.~\cite{Sirunyan:2018ruf} by comparing the vector LQ to the scalar LQ limits\footnote{Note that the bounds could be weakened in case the LQ is not purely a gauge boson by minimizing the cross section with respect to $\kappa_G$ and $\lambda_G$~\cite{Blumlein:1996qp,Blumlein:1998ym}}. Furthermore, the limits from PP differ for the various LQ representations~\cite{Diaz:2017lit}. In case of a small mass splitting among the $SU(2)_L$ components, as realized for $v\ll m,M$, their contributions add up to the total signal strength. This can be incorporated in the analysis by choosing an ``effective'' value of $\beta$ (originally parametrizing the branching fraction to electrons) which can then however be bigger than 1 (e.g. $\sqrt 2$ for $\lambda^{LR}_2$). Therefore, we extrapolated the $\beta$ dependence of the limits given in Refs.~\cite{Sirunyan:2018kzh,Sirunyan:2018btu} to account for these cases.

While the bounds from SP via {$qg \to {\rm LQ}\to \ell \ell q$} are quite weak~\cite{Khachatryan:2015qda,Mandal:2015vfa,Schmaltz:2018nls}, in case of first generation LQs much better bounds can be derived from SRP via $\ell q\to {\rm LQ}\to \ell q$~\cite{Buonocore:2020erb,Greljo:2020tgv} using the electron PDF of the proton~\cite{Buonocore:2020nai}. Since Ref.~\cite{Buonocore:2020erb} considers a simplified setup with $ue$ and $de$ interactions separately we have to adapt the limits for several of our LQ representations. First of all, as for PP, the small mass splitting between the $SU(2)_L$ components leads to overlapping signals (i.e. the cross sections of the components have to be added). In addition, we have to take into account the difference between the up and down quark PDFs, {which can be obtained for the relative strength of the $ue$ and $de$ limits given in Ref.~\cite{Buonocore:2020erb}}. Furthermore, if the LQ couples to a lepton doublet, we must adjust the branching ratio as it can decay to neutrinos whose signal is not included in the analysis. Finally, for VLQs we have to correct for the fact that, due to the Dirac algebra, the on-shell production cross section is \mbox{$\sigma_{\text{VLQ}}= 2\sigma_{\text{SLQ}}+\mathcal{O}(\alpha_{s})$} {for equal LQ couplings to fermions}.

Limits from DY-like signatures were derived in Ref.~\cite{Schmaltz:2018nls} based on the CMS search for resonant di-lepton pairs~\cite{Sirunyan:2018exx}, but they turn out to be less constraining than the bounds from SRP~\cite{Buonocore:2020erb}. Interestingly, the latest non-resonant di-lepton search of ATLAS{\footnote{{Note that in v1 and v2 of the ATLAS article a factor 2 in the definition of the Lagranigian in Eq. (1) was missing. We thank the ATLAS collaboration for confirming this.}}}~\cite{Aad:2020otl} can be used to obtain more stringent bounds\footnote{In principle also LEP bounds on ee-qq interactions~\cite{Schael:2013ita} could be used to constrain first generation LQs. Even though these limits can be directly applied for TeV scale LQs, they turn out to be weaker compared to LHC searches and low energy precision constraints.}. Here we have to take into account that Ref.~\cite{Aad:2020otl} assumed quark flavour universality which is not respected by most of the representations. This can be done by correcting for the fact that at $2\,$TeV the $uu\to \ell^+\ell^-$ cross section is a factor $\approx 1.7$ bigger than the $dd\to \ell^+\ell^-$ one for equal couplings. Furthermore, unlike for the analysis of Ref.~\cite{Schmaltz:2018nls} which is valid for low LQ masses, here care has to be taken if the LQ mass is within the LHC energy range. Following Ref.~\cite{Bessaa:2014jya}, the four-fermion approximation can be used if the LQ mass squared is bigger than four times the center of mass energy. As the highest energy used in the analysis of Ref.~\cite{Aad:2020otl} is $\approx 2\,$TeV, the limits can be applied for a LQ mass above $\approx 4\,$TeV~\cite{Bessaa:2014jya}. If the LQ is lighter, the limit is weakened. In particular for a LQ mass of $1\,$TeV the bound on the coupling is a factor $\approx 1.6$ ($\approx 2.1$) less stringent than extracted in the 4-fermion approximation~\cite{Bessaa:2014jya} in case of constructive (destructive) interference\footnote{For our numerical analysis we interpolated the points given in Fig.~2 of Ref.~\cite{Bessaa:2014jya} to estimate the correction factor.}.

\section{Phenomenological Analysis}
\label{sec:pheno}

In our phenomenological analysis we consider each LQ representation separately. In addition, we only allow for a single non-zero coupling at a time so that there are two scenarios for $\Phi_{1,2}$ and $V_{1,2}$ each. Therefore, we have fourteen scenarios in total with three free parameters in each case: the LQ mass ($m,M$), the coupling ($\lambda,\kappa$) and the angle $\beta$. The LHC limits and the bounds from parity violation are to a good approximation independent of $\beta$ (for $\beta=O(\theta_c)$). Here we will consider two cases, $\beta=0$ and $\beta=\theta_{c}$, corresponding to the down and up basis, respectively. While in the first case no effects in Kaon physics appear, bounds from $D^{0}-\bar{D}^{0}$ mixing are relevant for all LQ representations involving couplings to quark doublets. On the other hand, if $\beta=\theta_{c}$, no limits from $D$ physics can be obtained, but $K^+\to \pi^+ \nu\nu (e^+e^-)$ puts bounds on the parameter space.

\begin{figure*}
	\begin{center}
		\includegraphics[width=0.45\textwidth]{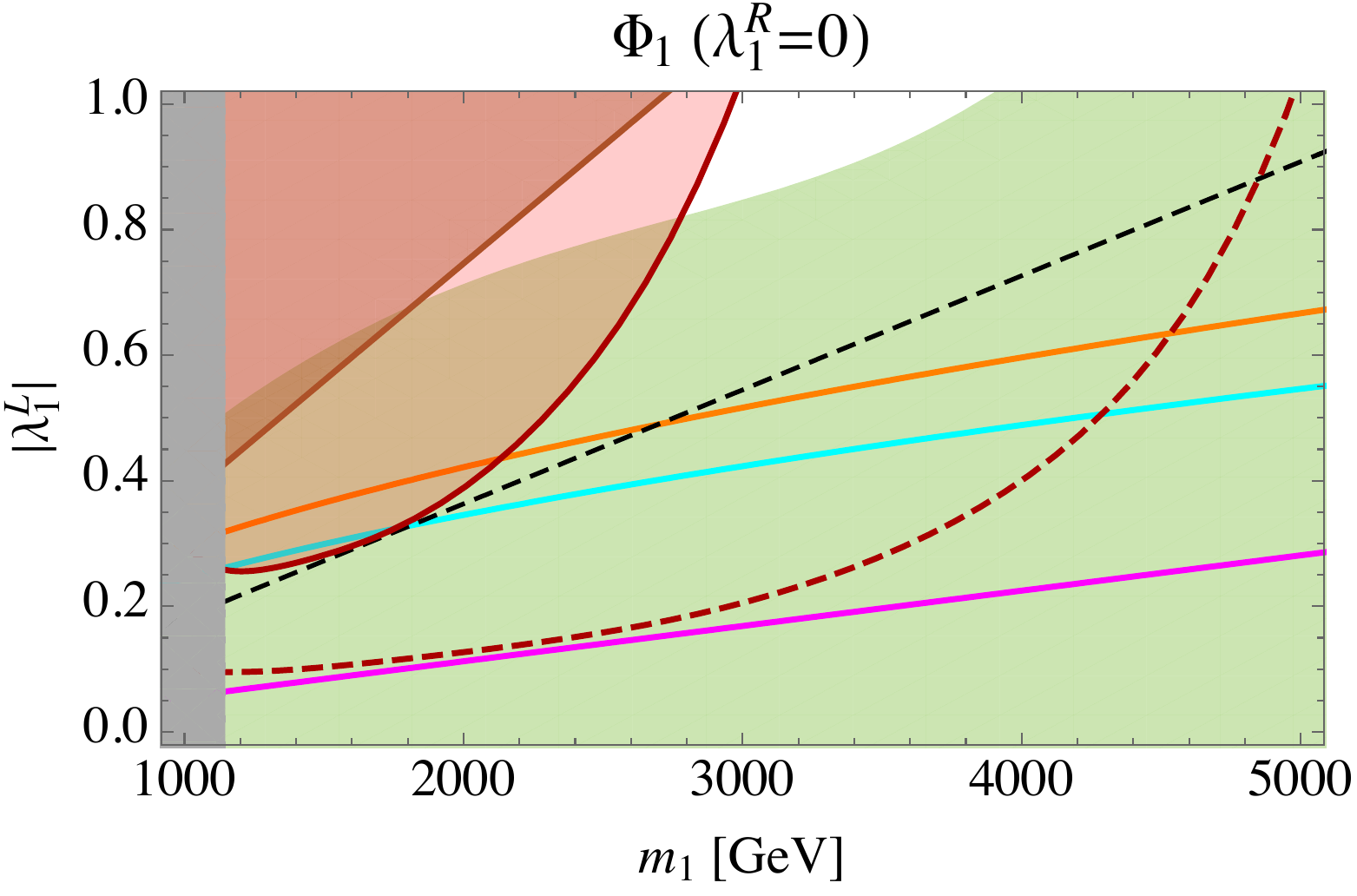}
		\quad
		\includegraphics[width=0.45\textwidth]{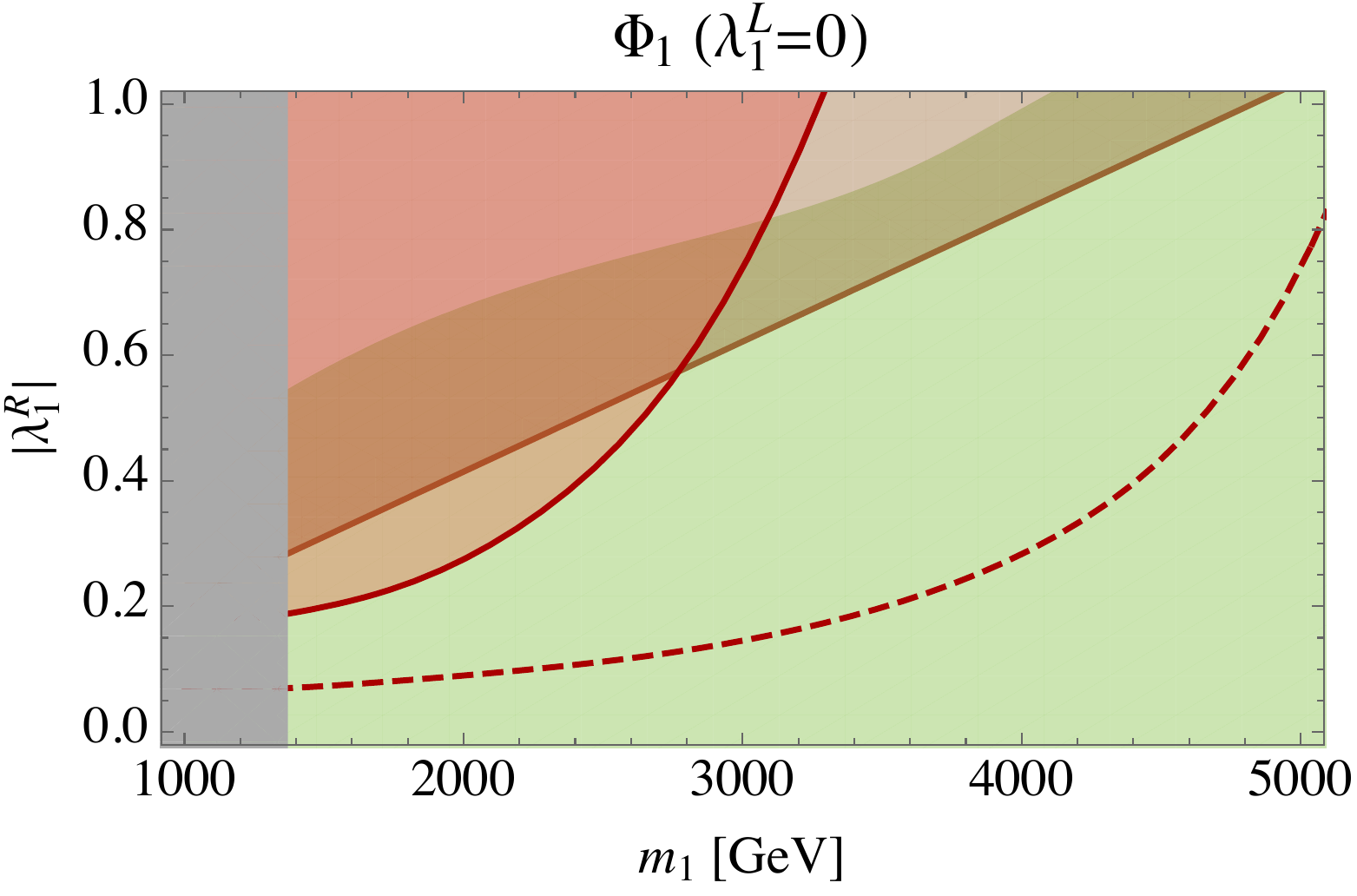}\\
		\includegraphics[width=0.45\textwidth]{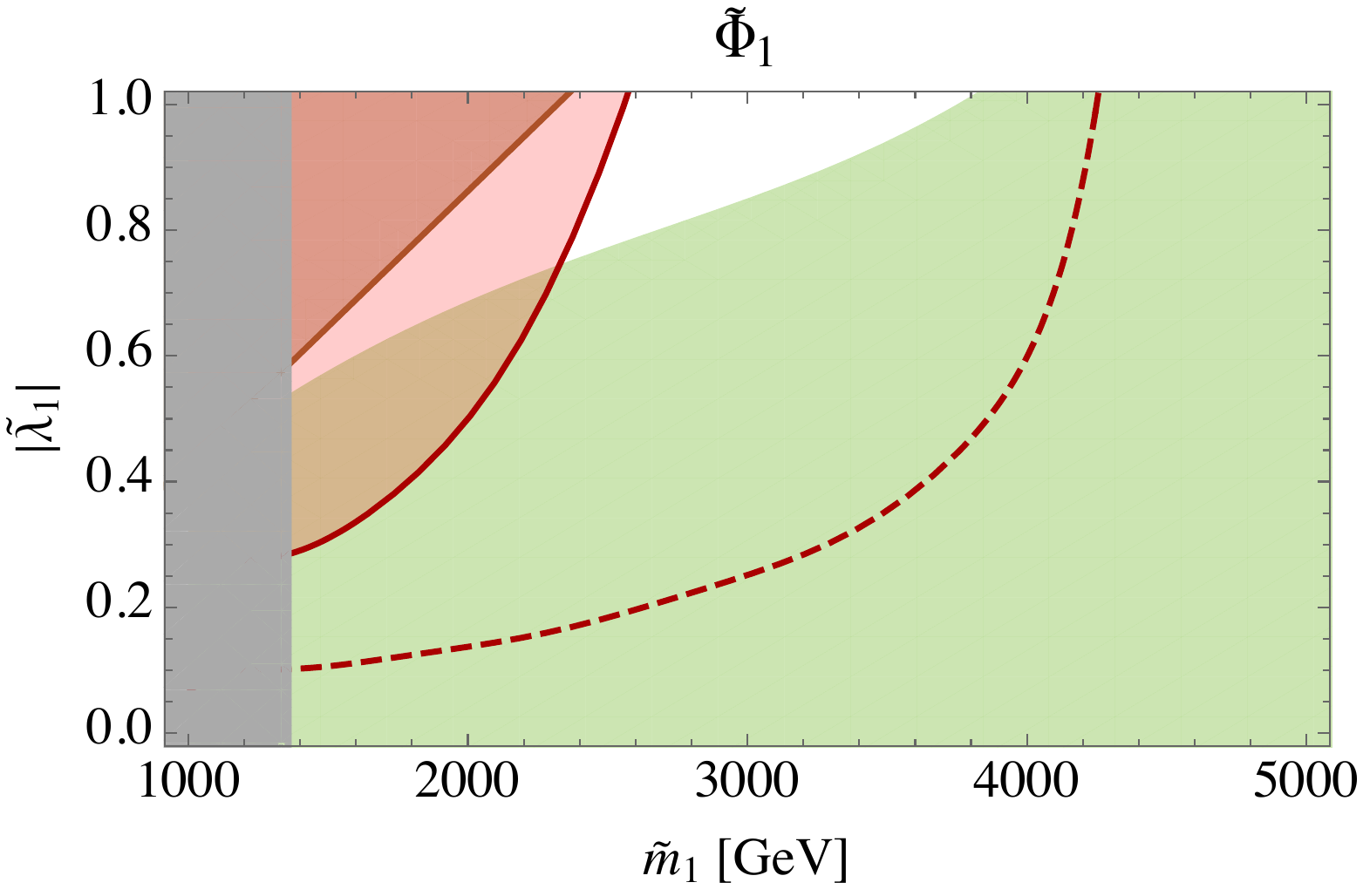}\qquad
		\includegraphics[width=0.45\textwidth]{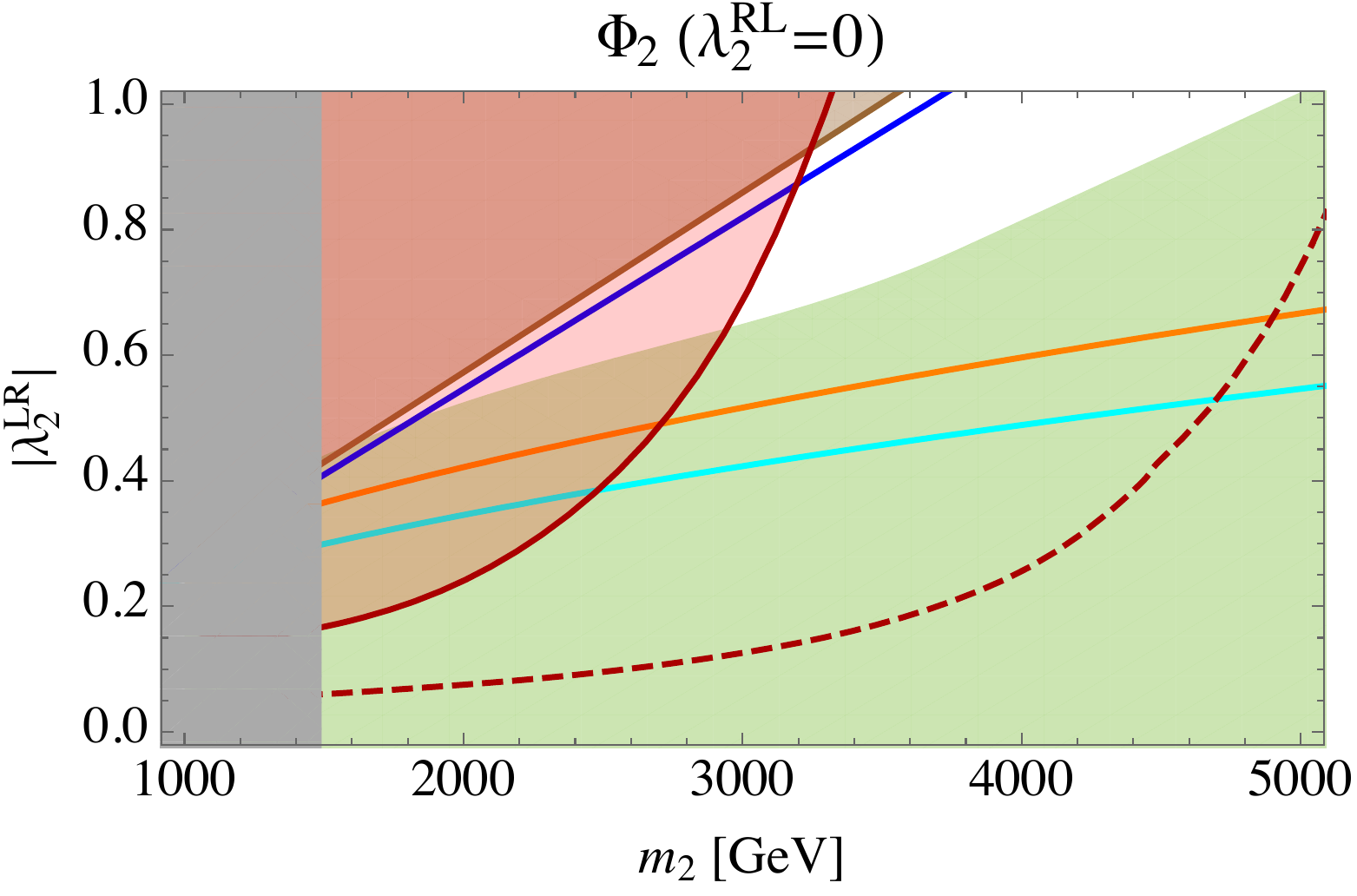}\\
		\includegraphics[width=0.45\textwidth]{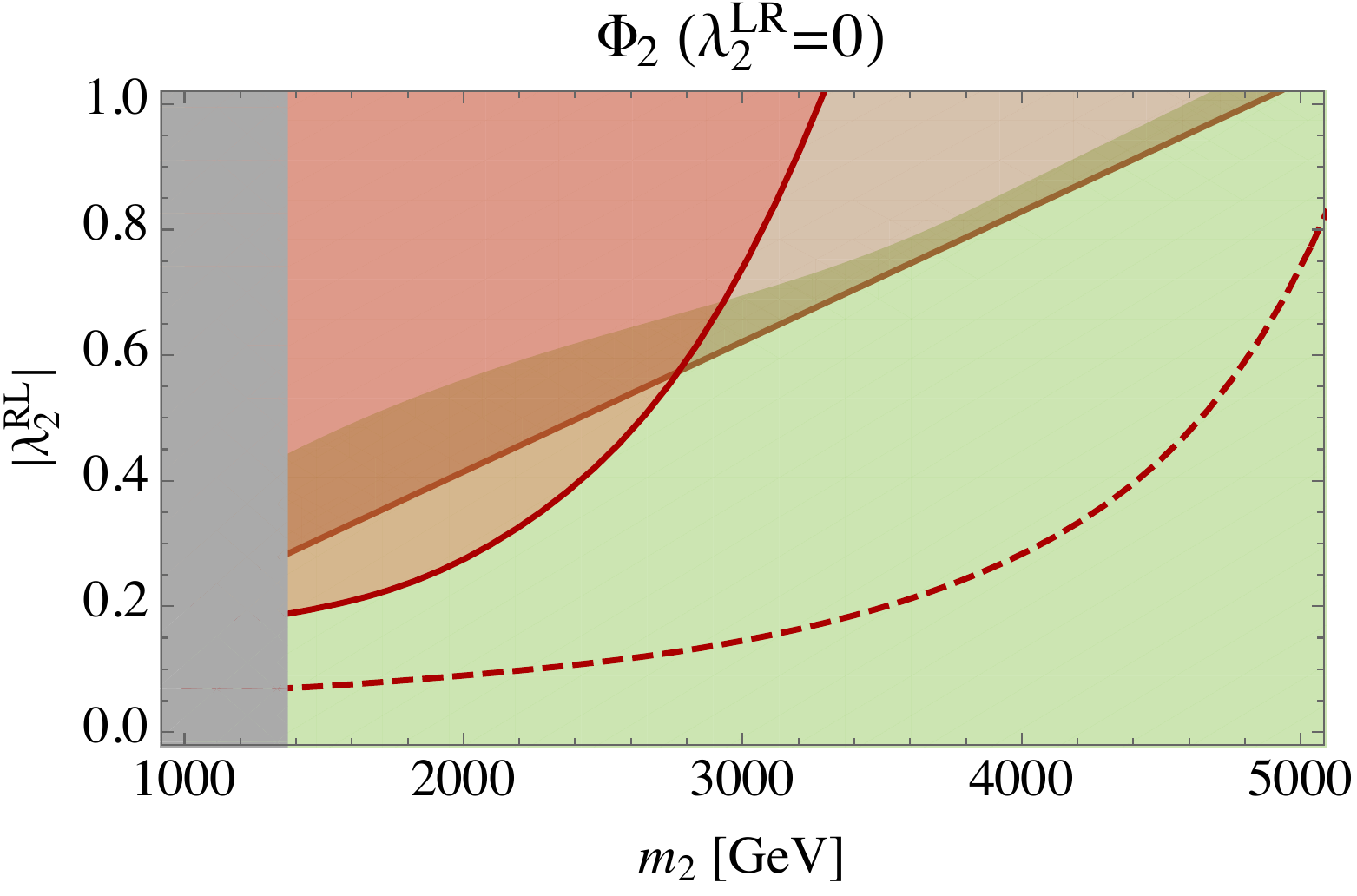}\qquad
		\includegraphics[width=0.45\textwidth]{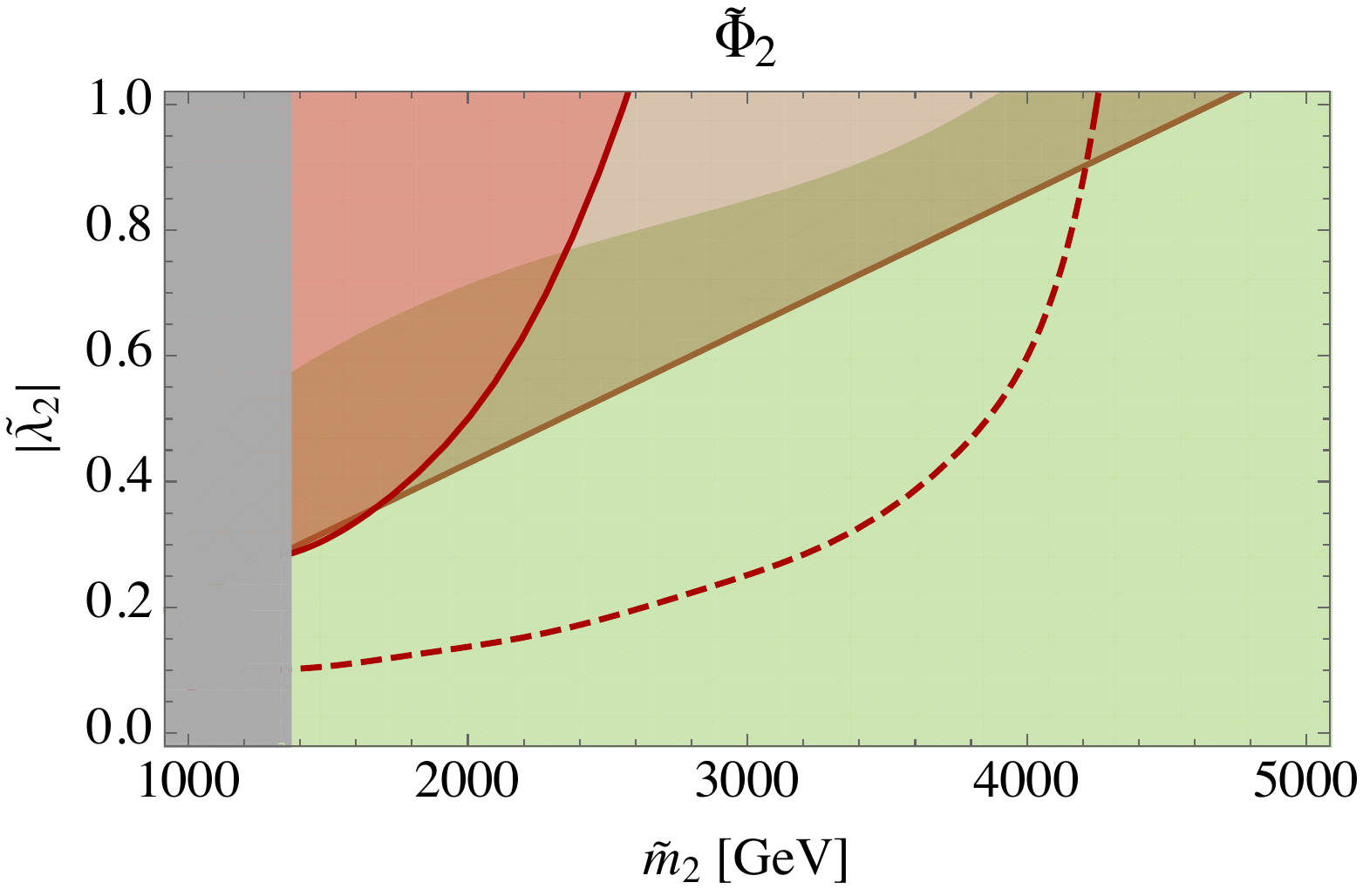}\\
		\includegraphics[width=0.7\textwidth]{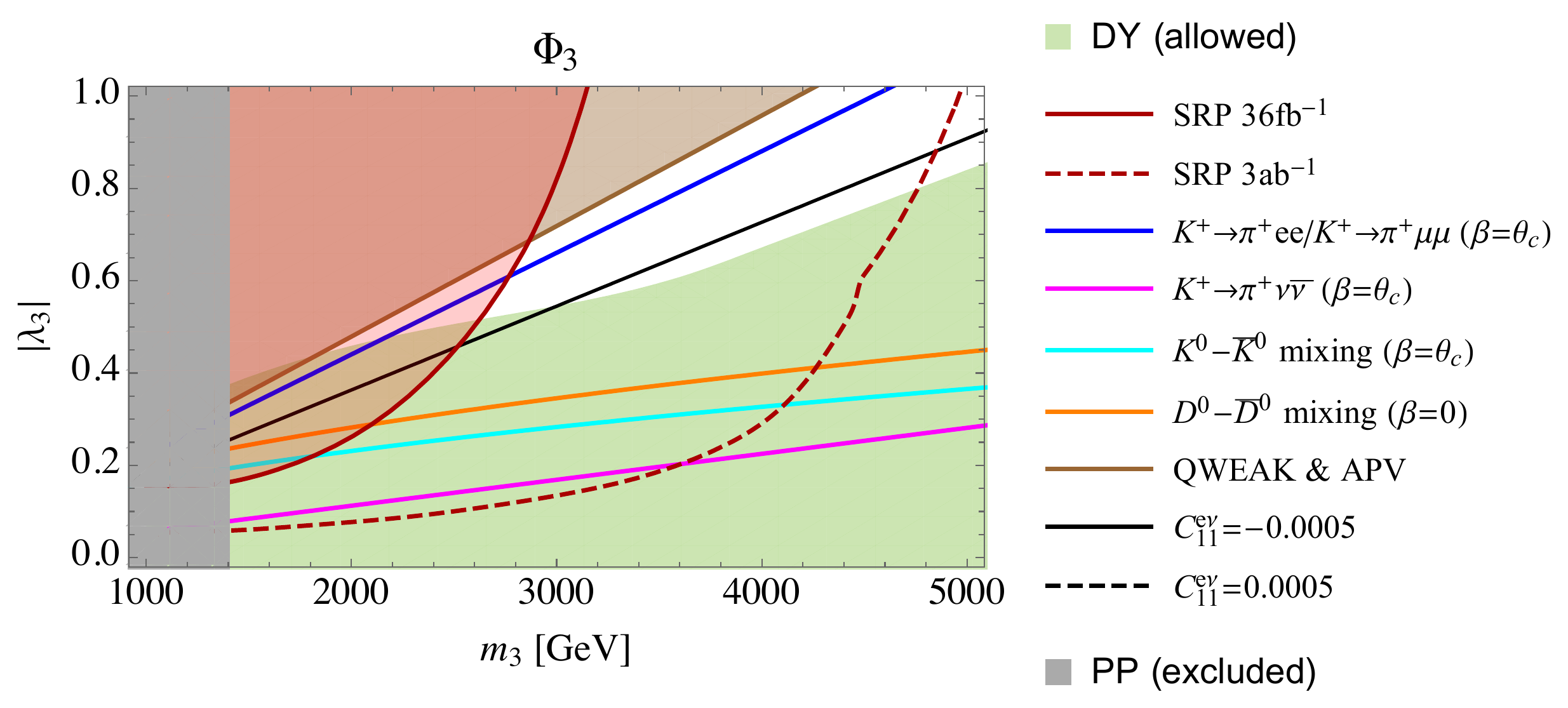}
	\end{center}
	\caption{Limits on the parameter space of first generation scalar LQs. The region above the colored lines is excluded. While LHC limits and the bounds from parity violation are to a good approximation independent of {$\beta$ (for $\beta= O(\theta_c)$)} the bounds from kaon and $D$ decays depend on it. We consider the two scenarios $\beta=\theta_{c}$ or $\beta=0$. In the first case, the kaon limits arise for LQ representations with left-handed quark fields while in the second case these limits are absent but bounds from $D^{0}-\bar{D}^{0}$ arise.}
	\label{fig:SLQ_plot}
\end{figure*}

\begin{figure*}
	\begin{center}
		\includegraphics[width=0.45\textwidth]{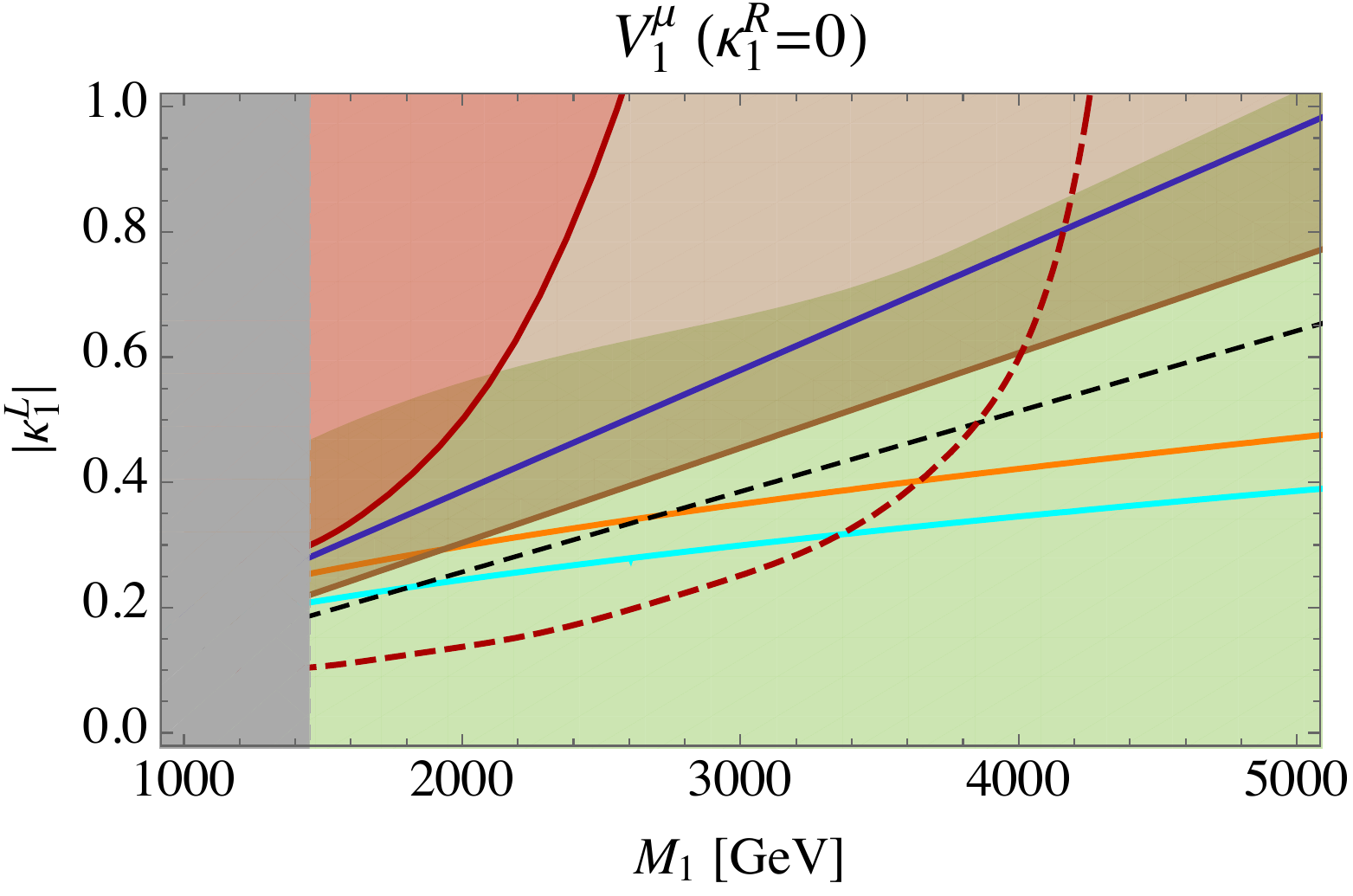}
		\quad
		\includegraphics[width=0.45\textwidth]{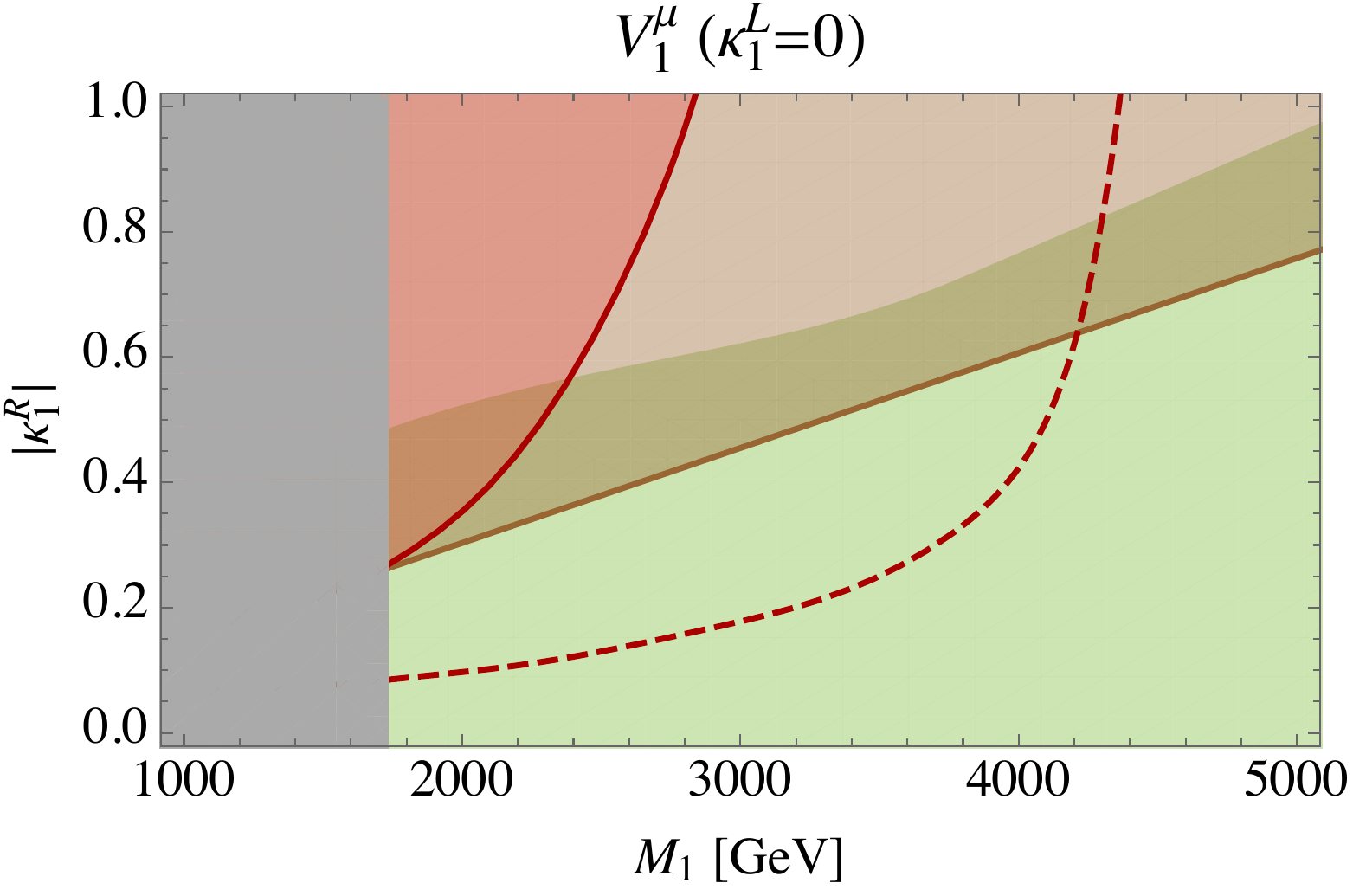}\\
		\includegraphics[width=0.45\textwidth]{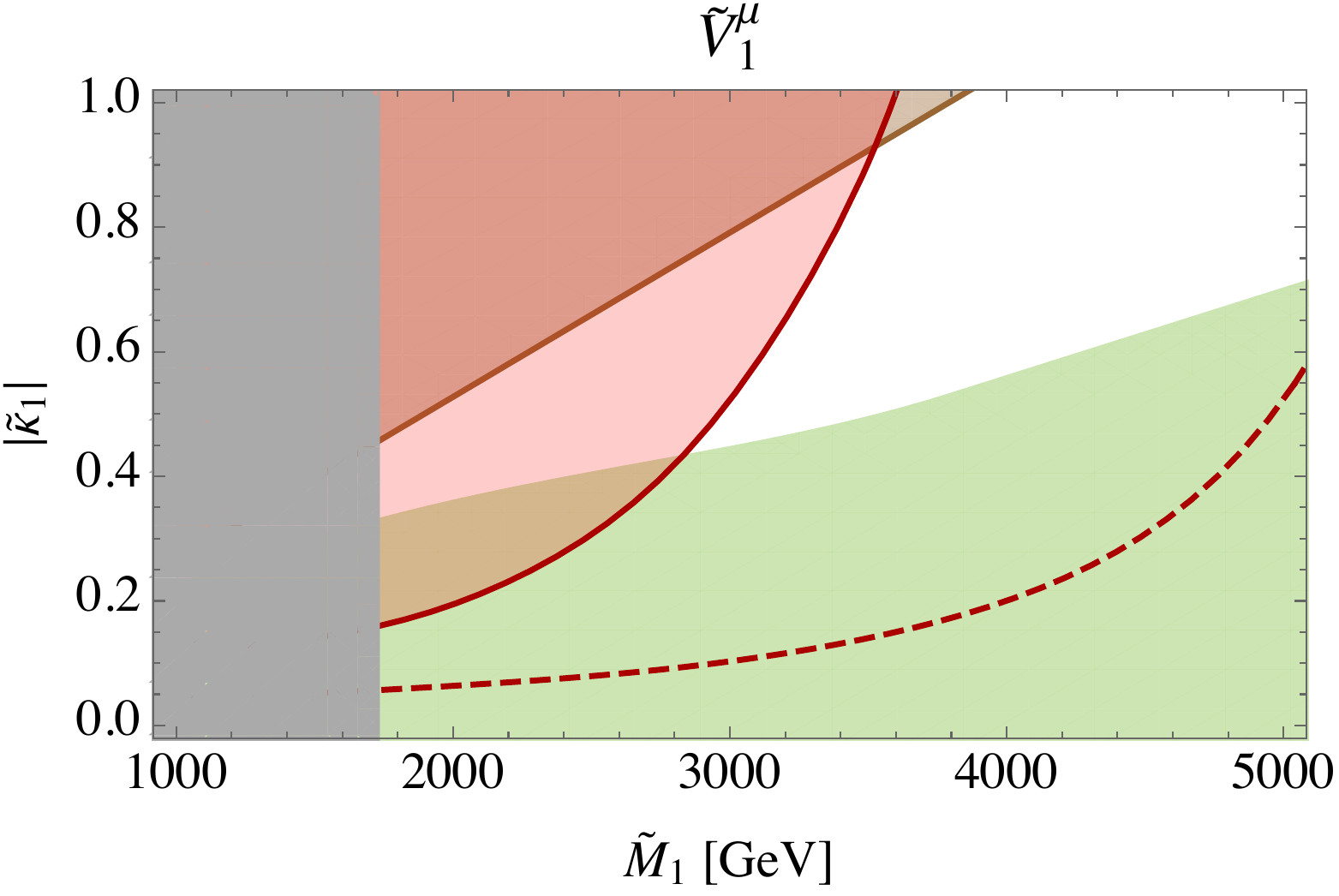}\qquad
		\includegraphics[width=0.45\textwidth]{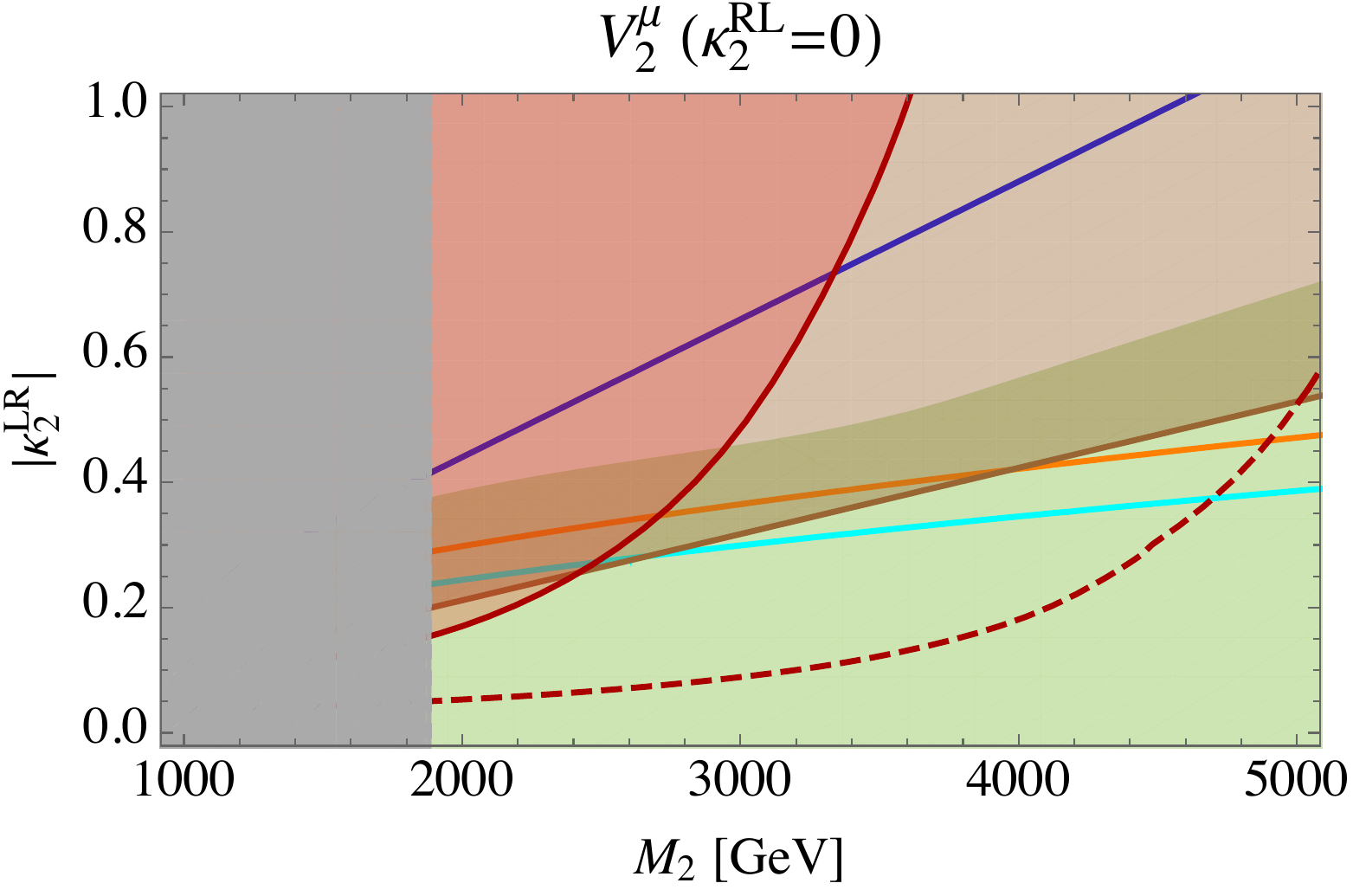}\\
		\includegraphics[width=0.45\textwidth]{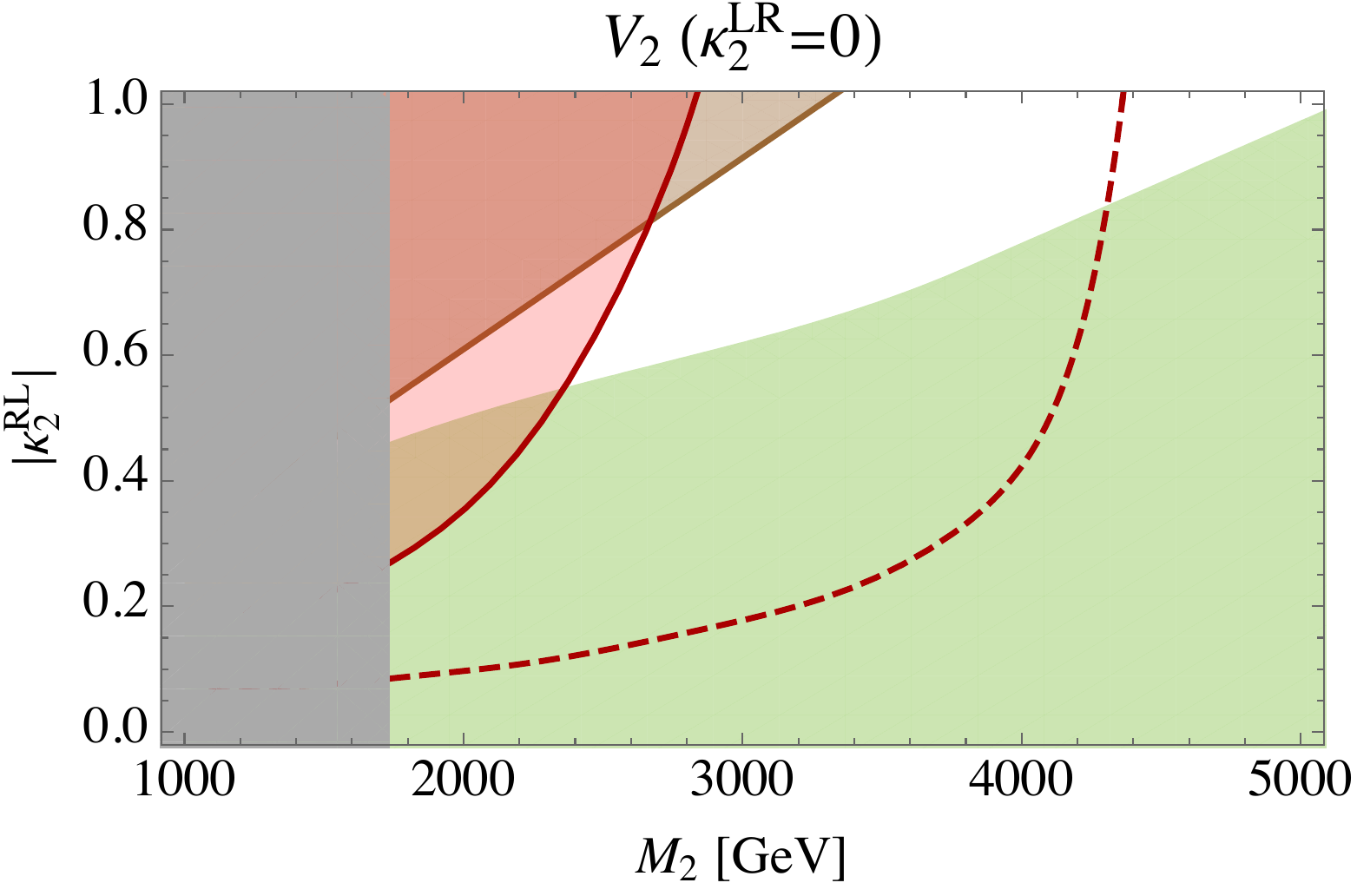}\qquad
		\includegraphics[width=0.45\textwidth]{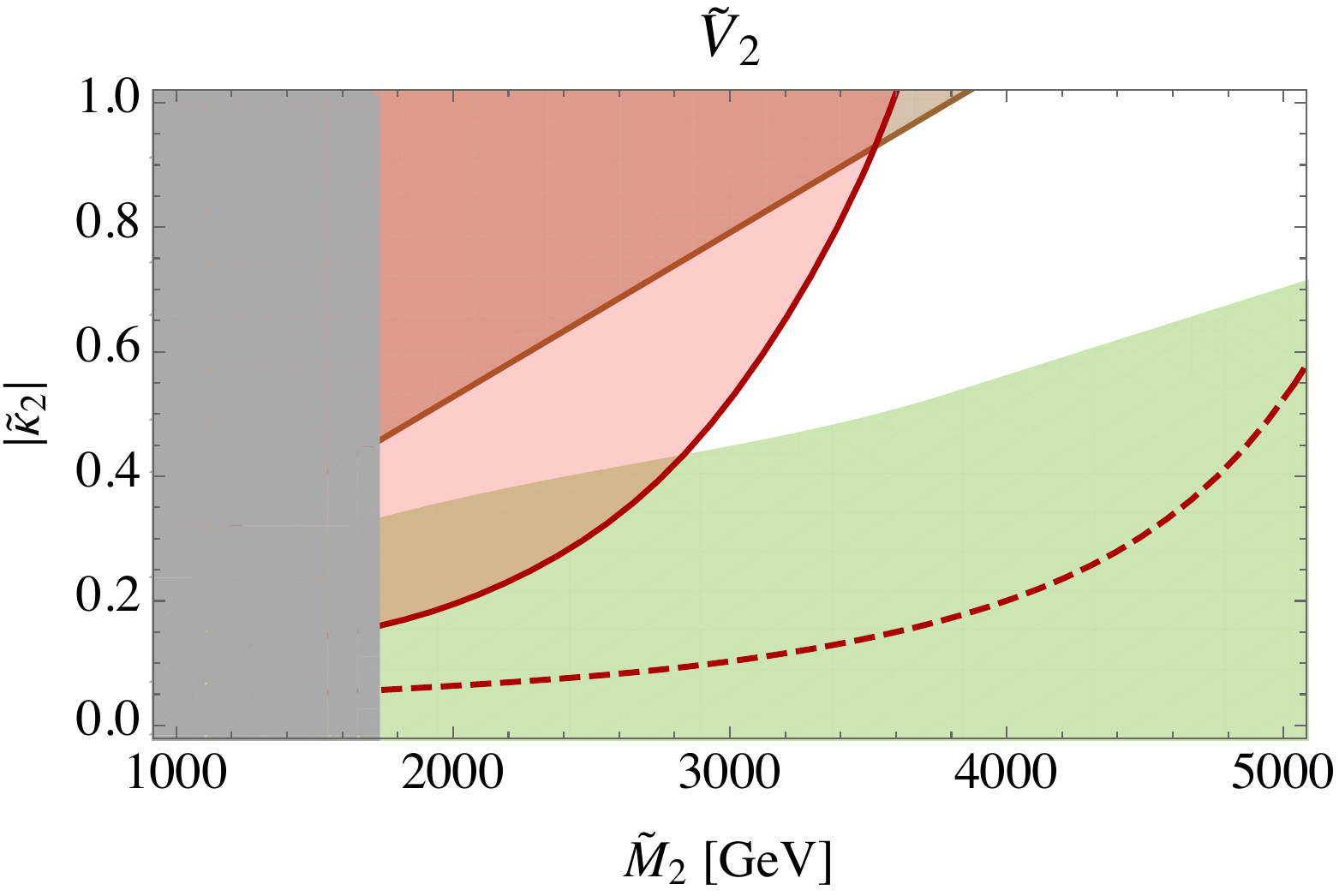}\\
		\includegraphics[width=0.7\textwidth]{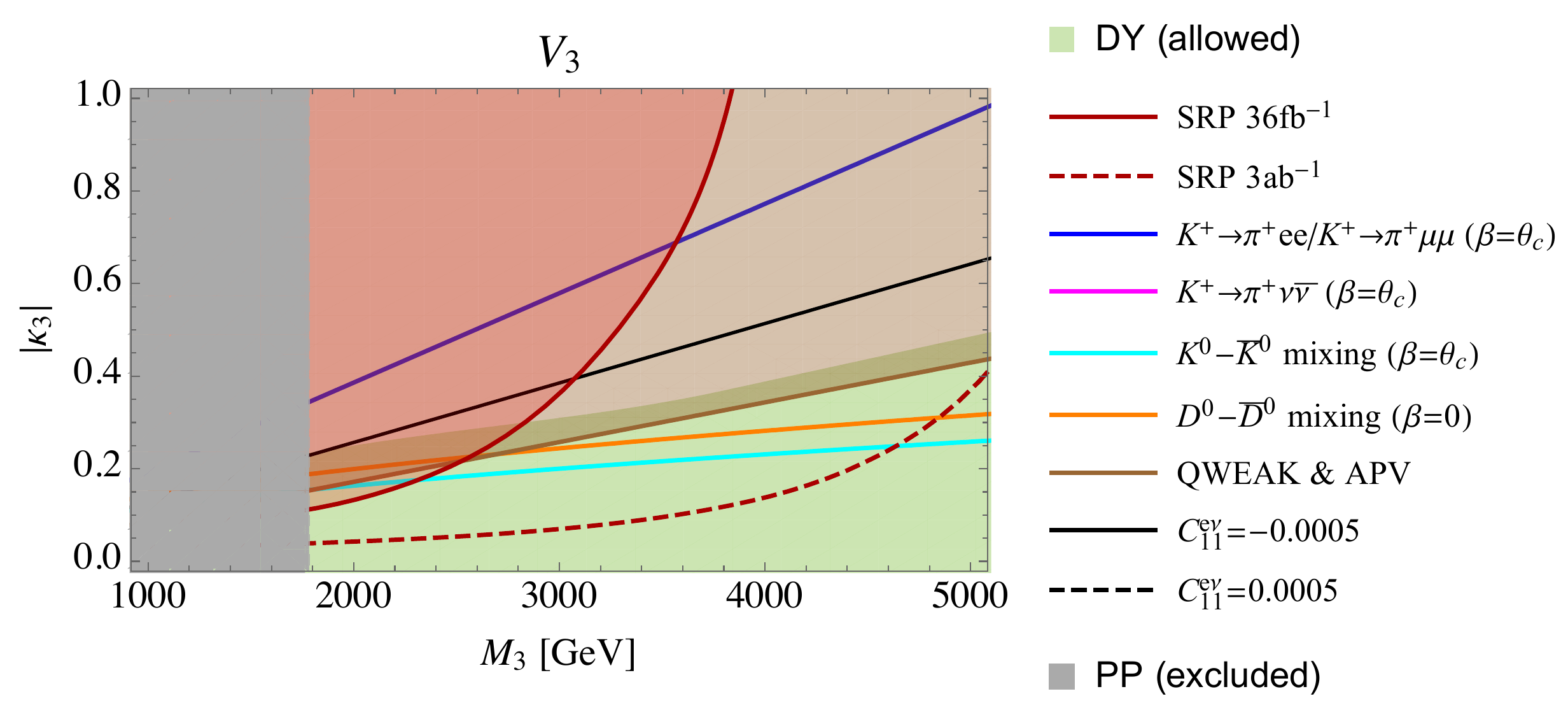}
	\end{center}
	\caption{Limits on the parameter space of first generation vector LQs. The region above the colored lines is excluded. While LHC limits and the bounds from parity violation are to a good approximation independent of {$\beta$ (for $\beta= O(\theta_c)$)} the bounds from kaon and $D$ decays depend on it. We consider the two scenarios $\beta=\theta_{c}$ or $\beta=0$. In the first case, the kaon limits arise for LQ representations with left-handed quark fields while in the second case these limits are absent but bounds from $D^{0}-\bar{D}^{0}$ arise.}
	\label{fig:VLQ_plot}
\end{figure*}

In Fig.~\ref{fig:SLQ_plot} and Fig.~\ref{fig:VLQ_plot} we show combined constraints (as well as the 3$ab^{-1}$ projection for SRP) on the parameter space of first generation LQs. All cases are constrained by LHC searches and parity violation experiments (QWEAK+APV) but bounds from Kaon and $D$ physics only appear in the case of couplings to quark doublets. In this case it is not possible to avoid both Kaon and $D$ bounds simultaneously, and the resulting limits are stringent. Furthermore, the ATLAS bounds on non-resonant di-lepton production are also very stringent and in fact more constraining than the DY bounds~\cite{Diaz:2017lit} (not displayed here) obtained from recasting resonant di-lepton searches~\cite{Sirunyan:2018exx}. Note that the 95\% CL limits from QWEAK+APS give quite different constraints on the various LQ representations since the central value is about $1\,\sigma$ off the SM prediction.

Only the representations $\Phi_{1,3}$ and $V_{1,3}$ generate a charged current whose strength is indicated by the black lines. Here the Cabibbo angle anomaly prefers negative values $C_{11}^{e\nu}\approx {-}0.001$. This disfavours $V_{1}$ ($\phi_{1}$) with $\lambda_1^L\neq0$ ($\kappa_1^L\neq0$) while it would in principle favour $V_3$ and $\Phi_3$. However, DY searches as well as $K^0-\bar K^0$ and/or $D^0-\bar D^0$ mixing exclude sizeable values of $C_{11}^{e\nu}$. Therefore, despite the fact that LQs can give tree-level effects in (super-allowed) beta decays, they cannot account for the deficit in first row CKM unitarity.

\section{Conclusion}
\label{conclusion}

In this article we performed a combined analysis of constraints on first generation LQs for all ten possible representations (five scalar and five vector ones). We included the constraints from parity violating experiments (QWEAK+APV) and LHC searches, in particular PP, SRP and DY searches. For the latter case, we find that the latest non-resonant di-lepton analysis of ATLAS provides stronger bounds than resonant searches recasted so far in the literature. As for left-handed quarks ''first generation`` can only be defined in the weak basis before EW symmetry breaking, unavoidable effects in Kaon and/or $D$ physics occur for the LQ representations coupling to quark doublets. Our results are depicted in Fig.~\ref{fig:SLQ_plot} and Fig.~\ref{fig:VLQ_plot} for scalar and vector LQs, respectively. One can see that all cases are constrained by parity violating experiments and LHC searches but only the cases which involve quark doublets are constrained by Kaon and/or $D$ physics. Furthermore, only 4 representations give rise to charged current effects where the Cabibbo angle anomaly prefers a destructive effect with respect to the SM. Such an effect can only be generated by $\Phi_3$ and $V_3$ and the possible size is too constrained by DY searches as well as $K^0-\bar K^0$ and/or $D^0-\bar D^0$ mixing to account for the anomaly.
\medskip

\begin{acknowledgements}
A.C. thanks Marc Montull for useful discussions. The work of A.C. is supported by a Professorship Grant (PP00P2\_176884) of the Swiss National Science Foundation. L.S. is supported by the ``Excellence Scholarship \& Opportunity Programme'' of the ETH Z\"urich.
\end{acknowledgements}

\bibliography{BIB}

\end{document}